\documentclass[prd,aps,showpacs,preprintnumbers,amssymb]{revtex4}

\usepackage{color}
\usepackage{epsf}

\def\e3p{$\eta \rightarrow 3 \pi$}

\begin{document}
\title{%
\hfill{\normalsize\vbox{%
\hbox{}
 }}\\
{Examining a possible cascade effect in chiral symmetry breaking}}

\author{Amir H. Fariborz
$^{\it \bf a}$~\footnote[1]{Email:
 fariboa@sunyit.edu}}

\author{Renata Jora
$^{\it \bf b}$~\footnote[2]{Email:
 rjora@theory.nipne.ro}}

\affiliation{$^{\bf \it a}$ Department of Matemathics/Physics, SUNY Polytechnic Institute, Utica, NY 13502, USA}
\affiliation{$^{\bf \it b}$ National Institute of Physics and Nuclear Engineering PO Box MG-6, Bucharest-Magurele, Romania}

\date{\today}

\begin{abstract}
We examine a toy model and a cascade effect for confinement and  chiral symmetry breaking which consists in several phase transitions corresponding to the formation of bound states and chiral condensates with different number of fermions for a strong group. We analyze two examples: regular QCD where we calculate the "four quark" vacuum condensate and a preon composite model based on QCD at higher scales.  In this context we also determine the number of flavors at which the second chiral and confinement phase transitions occur and discuss the consequences.

\end{abstract}
\maketitle

\section{Introduction}
A general description of the phase diagram of QCD includes several phases among which confinement and chiral symmetry breaking are the most relevant for the low energy regime. However what it is known and can be extracted from experiments regarding the low lying meson spectroscopy  suggest that in order to explain thier properties one would need to take into account besides the usual "two quark" mesons also the possibility that "four quark" states might also exist. Tetraquark states were first introduced by Jaffe \cite{Jaffe} in the MIT bag model and later explored in several studies \cite{Jora7}-\cite{Polosa3}. It was shown that the unusual inverted scalar spectrum may be determined by a large "four quark" composition. If bound states of  four quarks exist then it is natural to consider also their condensates. The appearance of  multi fermion bound states and condensates
is not taken usually into account as a separate phenomenon in drawing the phase diagram of a strong theory because it is assumed that the confining and chiral symmetry breaking processes are somehow continuous and do not yield multiple phase transitions.

 In this paper we examine a picture where the actual confinement and chiral symmetry breaking take place in steps or stages each one corresponding to a different phase transition.  We shall call this process " cascade confinement and chiral symmetry breaking" and in essence consists in  a series of phase transitions each one occurring at a different scale and coupling constant and corresponding to a different number of fermions that bound together or condensate. We will study in detail two cases: QCD at a regular scale where quark and bound states of quark exist; QCD at a higher scale where the elementary fermions are the preons that constitute the quark components, composite model proposed in \cite{Jora1}.  The possibility that the formation of "four quark" condensate represents a separate phase transition was recently introduced in  \cite{Pisarski}.

 Section II contains the set-up and the four beta functions that will be of interest in this work. In section III we calculate the "four quark" condensate in a simple Nambu Jona Lasinio mechanism. In section IV we discuss the phenomenon of cascade confinement and chiral symmetry breaking  for regular QCD whereas in section V we analyze the same issues for QCD at higher scales that contains preons in a composite picture. Section VI is dedicated to conclusions.

\section{The set-up}
In this work we consider a picture in which a nonabelian gauge group may present the phenomenon of cascade confinement and chiral symmetry breaking, indicating the possibility that confinement and chiral symmetry breaking
may take place in steps or stages as one goes to lower scales where the coupling constant gets larger and multiparticle states or condensates may form. We shall discuss in particular the $SU(3)$ group but our findings may extend to other strong theories. We assume that at a high scale the fermion fields situated in some representation $R$ of the gauge group confine and form singlet  bound states of two or three fermions. At a slightly lower scale the two fermion condensates appear and some chiral symmetry breaking occurs. At similar scales slightly bound two or three fermion states form situated in a representation of the gauge group that  is lower or equal in dimension with the dimension of $R$. We call this the breaking of representation to lower ones. Such states modify the initial beta function of the gauge group and lead to a different behavior of the strong coupling constant. At an even lower scale a second confinement and chiral symmetry breaking may take place corresponding to the formation of four fermion or five fermion singlets and that of the possible tetrafermion condensates.

First we will need the beta function for an $SU(N)$ gauge theory with fermion and scalars in an arbitrary representation \cite{Machacek}-\cite{Tarasov}:
\begin{eqnarray}
\frac{d a}{d \ln(\mu^2)}&=&\beta(a)=-\beta_0a^2-\beta_1a^3-...
\nonumber\\
\beta(a)&=&-a^2[\frac{11}{3}C_2(G)-\frac{4}{3}S_2(F)N_f-\frac{1}{6}S_2(S)N_s]-
\nonumber\\
&&a^3[\frac{34}{3}C_2(G)^2-4C_2(F)S_2(F)N_f-\frac{20}{3}C_2(G)S_2(F)N_f-2C_2(S)S_2(S)N_s-\frac{1}{3}C_2(G)S_2(S)N_s],
\label{betafisrst6453738i}
\end{eqnarray}
where $a =\frac{g^2}{16\pi^2}$, $N_f$ is the number of four component fermions, $N_S$ is the number of real scalar degrees of freedom, $G$ is the adjoint representation, $F$ is the fermions representation and $S$ is the scalar one.
Moreover $C_2(R)$ is the quadratic Casimir operator for the representation $R$, $C_2(R)=T^A_RT^A_R$ where $T^A_R$ are the generators in the representation $R$ and  $S_2(R)$ is the Dynkin index for the same representation such that:
\begin{eqnarray}
{\rm Tr}_R(T^A_RT^B_R)=S_2(R)\delta^{AB}.
\label{index53758}
\end{eqnarray}
We further use
\begin{eqnarray}
S_2(R_1\times R_2)=S_2(R_1)D(R_2)+S_2(R_2)D_(R_1)
\nonumber\\
D(R_1\times R_2)C_2(R_1 \times R_2)=D(G)S_2(R_1 \times R_2),
\label{res6295748}
\end{eqnarray}
where $D(R)$ is the dimension of the representation $R$.
We are interested in the indices corresponding to the fundamental, adjoint, antisymmetric and product representations (Note that the conjugate representations will have the same indices).
\begin{eqnarray}
&&S_2(N)=\frac{1}{2}\,\,\,\,\,\,\,\,\,\,C_2(N)=\frac{N^2-1}{2N}
\nonumber\\
&&S_2(G)=N\,\,\,\,\,\,\,\,\,\,C_2(G)=N
\nonumber\\
&&S_2(antisym)=\frac{N-2}{2}\,\,\,\,\,\,\,\,\,\,C_2(antisym)=N-\frac{2}{N}-1
\nonumber\\
&&S_2(N\times N)=N\,\,\,\,\,\,\,\,\,\,C_2(N \times N)=\frac{N^2-1}{N},
\label{res74536728}
\end{eqnarray}
where $N$ in the bracket corresponds to the fundamental representation, $G$ to the adjoint representation, $antisym$ to the antisymmetric one and $N \times N$ to the product representation.

There are two particular cases that we shall discuss in the present work.  The first one is regular QCD with three light flavors. In the initial gauge group at high energies the beta function is given by:
\begin{eqnarray}
&&\beta_0=\frac{11}{3}N-\frac{2}{3}N_f
\nonumber\\
&&\beta_1=\frac{34}{3}N^2-2\frac{N^2-1}{2N}N_f-\frac{10}{3}NN_f
\label{coef453789}
\end{eqnarray}
and throughout this work we will consider only the first two coefficients that are renormalization scheme independent. Here $N$ is the number of colors whereas $N_f$ is the number of flavors.

At a lower scale diquark and baryon like states situated in an antisymmetric and fundamental representations of $SU(3)$ form. The number of real scalar degrees of freedom and baryons that can appear where we considered only the low lying spin $\frac{1}{2}$ baryons is $2N_f^2$ and $N_f^2\frac{N_f-1}{2}$. Here we took  into account the fact that $N_f$ fermions lead to the formation of $N_f^2$ scalar states and $N_f^2$ pseudoscalar states which correspond to $2N_f^2$ real scalar degrees of freedom. For the number of baryons we consider  that in the structure $\bar{\Psi}\chi\tau$ where each entry corresponds to a fermion the last two states must be antisymmetric and thus different in order to form a baryon with the spin $\frac{1}{2}$. Thus the total number of states is $N_f(\frac{N_f(N_f-1)}{2})$ where $(\frac{N_f(N_f-1)}{2})$ corresponds to the number of possibilities for the last two entries (see \cite{Jora1} for details about the counting of states).  Then the  absence of the initial colored fermion states and the formation  of baryon and meson like colored states modify the beta function ($\beta'(a)=-\beta_0'a^{2}-\beta_1'a^{3}$)  according to:
\begin{eqnarray}
&&\beta_0'=\frac{11}{3}N-\frac{2}{3}N_f^2\frac{N_f-1}{2}-\frac{1}{6}N_f^2(N-2)
\nonumber\\
&&\beta_1'=\frac{34}{3}N^2-\frac{N^2-1}{N}N_f^2\frac{N_f-1}{2}-\frac{10}{3}NN_f^2\frac{N_f-1}{2}-2(N-\frac{2}{N}-1)\frac{N-2}{2}2N_f^2-\frac{1}{3}N\frac{N-2}{2}2N_f^2.
\label{functr6645}
\end{eqnarray}

On the other hand we will consider a model proposed in \cite{Jora1} where at a higher scale the $SU(3)$ group contains $N_{f}'$ fermions in the complex conjugate product representation and study the possibility of cascade confinement down to QCD. At a higher scale the beta function is given by $\beta''(a)=-\beta_0''a^{\prime2}-\beta_1''a^{\prime3}$ where:
\begin{eqnarray}
&&\beta_0''=\frac{11}{3}N-\frac{4}{3}NN_{f}'
\nonumber\\
&&\beta_1''=\frac{34}{3}N^2-4\frac{N^2-1}{N}NN_{f}'-\frac{20}{3}N^2N_{f}'.
\label{thirdbetafunc}
\end{eqnarray}
The second confinement and chiral symmetry breaking transition occurs according to a new beta function that contains three and two preon states situated in the $3$ or $3^*$ representation. The number of slightly bound three preon states is $3N_{f}^{\prime2}\frac{N_{f}'-1}{2}$ where we took into account the three possible bindings  whereas that of two preon scalars is $4N_{f}^{\prime2}$ where there are two possible bindings.  The new beta function is
$\beta'''(a)=-\beta_0'''a^2-\beta_1'''a^3$ where:
\begin{eqnarray}
&&\beta_0'''=\frac{11}{3}N-\frac{2}{3}3N_{f}^{\prime2}\frac{N_{f}'-1}{2}-\frac{1}{6}\frac{N-1}{2}4N_{f}^{\prime2}
\nonumber\\
&&\beta_1'''=\frac{34}{3}N^2-\frac{N^2-1}{N}3N_{f}^{\prime2}\frac{N_{f}'-1}{2}-\frac{10}{3}N3N_{f}^{\prime2}\frac{N_f'-1}{2}-
2(N-\frac{2}{N}-1)\frac{N-2}{2}4N_f^{\prime2}-\frac{1}{3}N\frac{N-2}{2}4N_f^{\prime 2}.
\label{res528901}
\end{eqnarray}

\section{An estimate of the tetraquark condensate}

Starting from the premises that the possibility of a cascade-type  confinement down to QCD occurs we will estimate the tetraquark condensate in a Nambu Jona Lasinio approach from first principles. According to our picture at some scale and for the anomalous dimension of the quark mass operator $\gamma_m=1$ and $\beta(a)=0$ the quark-antiquark vacuum condensate appears.  Here $a=\frac{g^2}{16\pi^2}$ and $g$ is the strong coupling constant. At a lower scale the coupling is larger and slightly bound diquark or three quark states appear in the color triplet or antitriplet representations.  These states will modify the beta function and at an even lower scale will lead to the formation of "four quark" condensates.

We consider QCD with three light quark flavors and three colors and assume that at smaller coupling and thus higher scale the gluon field will gain a mass $m_A$ such that the gluon fields can be integrated out for scales lower than this to produce an effective Nambu Jona Lasinio type model. The initial vertex of interest is,
\begin{eqnarray}
ig\bar{\Psi}^A_i\gamma^{\mu}(t^a)_{AB}\Psi_j^BA^a_{\mu},
\label{vert4554}
\end{eqnarray}
where $A$ and $B$ are color indices and $i$ and $j$ are flavor ones. The four quark interaction term is then extracted from the square of the term  in Eq. (\ref{vert4554}) in the functional approach:
\begin{eqnarray}
&&\frac{i}{2}\int d^4 x d^4 y \bar{\Psi}^A_i(x)\gamma^{\mu}(t^a)_{AB}\Psi^B_i(x)\bar{\Psi}^C_j(y)\gamma^{\nu}(t^b)_{CD}\Psi^D_j(y)\delta_{ab}\int \frac{d^4k}{(2\pi)^4}(g_{\mu\nu}-\frac{k_{\mu}k_{\nu}}{k^2})
\frac{1}{k^2-m_A^2}\exp[-ik(x-y)]
\nonumber\\
&&\approx-i\frac{3}{8m_A^2}\int d^4 x \bar{\Psi}^A_i(x)\gamma^{\mu}(t^a)_{AB}\Psi^B_i(x)\bar{\Psi}^C_j(x)\gamma^{\nu}(t^b)_{CD}\Psi^D_j(x).
\label{res54664}
\end{eqnarray}
Here we approximated the propagator to be equal to the inverse squared mass of the gluon field and for the term proportional to $\frac{k_{\mu}k_{\nu}}{k^2}$ we used the expansion in the gamma matrices basis and two Fierz transformation that took into account only the possible scalar contributions. We then further use,
\begin{eqnarray}
(t^a)_{AB}(t^a)_{CD}=\frac{1}{2}[\delta_{AD}\delta_{BC}-\frac{1}{3}\delta_{AB}\delta_{CD}]
\label{gr66575}
\end{eqnarray}
and also the Fierz transformation,
\begin{eqnarray}
\bar{\Psi}_1\gamma^{\mu}\Psi_2\bar{\Psi}_3\gamma^{\mu}\Psi_4=-\frac{1}{4}[4\bar{\Psi}_1\Psi_4\bar{\Psi}_3\Psi_2-
2\bar{\Psi}_1\gamma^{\mu}\Psi_4\bar{\Psi}_3\gamma^{\mu}\Psi_2-2\bar{\Psi}_1\gamma^{\mu}\gamma^5\Psi_4\bar{\Psi}_3\gamma^{\mu}\gamma^5\Psi_2-4\bar{\Psi}_1\gamma^5\Psi_4\bar{\Psi}_3\gamma^{5}\Psi_2],
\label{fierz54664}
\end{eqnarray}
to determine the corresponding scalar contribution:
\begin{eqnarray}
i\frac{3}{16m_A^2}\bar{\Psi}^A_i\Psi^A_j\bar{\Psi}^C_j\Psi^C_i.
\label{res65748996}
\end{eqnarray}
Using the equation of motion to extract the vacuum condensate we obtain:
\begin{eqnarray}
i\gamma^{\mu}\partial_{\mu}\Psi^A_j+i\frac{3}{8m_A^2}g^2\Psi^A_j\langle\bar{\Psi}^C_j\Psi^C_j\rangle+...=0,
\label{mot77686}
\end{eqnarray}
which leads to:
\begin{eqnarray}
m_q=-\frac{3}{8m_A^2}g^2\langle\bar{\Psi}^C_j\Psi^C_j\rangle=\frac{3}{4}g^2\alpha,
\label{res55464}
\end{eqnarray}
where we denoted the scalar vacuum expectation value as \cite{Jora2}:
\begin{eqnarray}
\alpha=-\frac{1}{2m_A^2}\langle\bar{\Psi}^C_j\Psi^C_j\rangle.
\label{m7766579}
\end{eqnarray}
Note that in the quark condensate there is summation over the number of colors but the flavor is fixed and we work in the $SU(3)$ invariant limit.

Next step is to find  a four scalar interaction term suitable for the diquark states. These are situated in an antitriplet of color and antitriplet of flavor according to the structure \cite{Jora3}:
\begin{eqnarray}
&&L^{gE}=\epsilon^{gab}\epsilon^{EAB}q^T_{aA}C^{-1}(\frac{1+\gamma^5}{2})q_{bB}
\nonumber\\
&&R^{gE}=\epsilon^{gab}\epsilon^{EAB}q^T_{aA}C^{-1}(\frac{1-\gamma^5}{2})q_{bB}.
\label{diq6645758}
\end{eqnarray}
We assume that the diquark states interact with the gauge fields as usual but with a different strong coupling constant $g'$ that runs with the new beta function stated in Eq. (\ref{functr6645}) that contains diquark scalars and triplet baryons.

The interaction term in the Lagrangian is:
\begin{eqnarray}
g^{\prime 2}L^{gE \dagger}L^{gF}(t^a)_{EB}(t^b)_{BF}A^a_{\mu}A^{b\mu},
\label{int45546}
\end{eqnarray}
with a similar term corresponding to the right handed states.
Then the partition function leads to the following four scalar interaction term:
\begin{eqnarray}
&&-g^{\prime 4}\int d^4x L^{gE \dagger}(x)L^{gF}(x)(t^a)_{EB}(t^b)_{BF}A^a_{\mu}(x)A^{b\mu}(x) d^4x\times
\nonumber\\
&&\int d^4y R^{tM\dagger}(y)R^{tN}(y)(t^d)_{MP}(t^e)_{PN}A^d_{\nu}(y)A^{e\nu}(y)\Rightarrow
\nonumber\\
&&3g^{\prime 4}\Bigg[\frac{1}{4}[(N-\frac{4}{N})\delta_{EN}\delta_{MF}+(1+\frac{2}{N^2})\delta_{EF}\delta_{MN}]\Bigg]\int \frac{d^4k}{(2\pi)^4}\frac{1}{(k^2-m_A^2)^2}+...
\label{actter55466}
\end{eqnarray}
where we extracted only the local interaction. Since we are interested only in contributions that lead to tetraquark condensate we can further process Eq. (\ref{actter55466}) to obtain;
\begin{eqnarray}
&&B=i 3g^{\prime 4}[\frac{1}{4}(N-\frac{4}{N})]\frac{1}{16\pi^2}\int d(k_E^2) \frac{k_E^2}{(k_E^2+m_A^2)^2}\times
\nonumber\\
&&\int d^4 x L^{gE \dagger}(x)R^{tE}(x)R^{tF \dagger}(x)L^{gF}(x).
\label{filre619047}
\end{eqnarray}

Clearly we need to evaluate the integral:
\begin{eqnarray}
I=\frac{1}{16\pi^2}\int d(k_E^2) \frac{k_E^2}{(k_E^2+m_A^2)^2}=\frac{1}{16\pi^2}\Bigg[\ln[1+\frac{\Lambda^2}{m_A^2}]-\frac{\Lambda^2}{\Lambda^2+m_A^2}\Bigg],
\label{int65748}
\end{eqnarray}
where $\Lambda$ is the cut-off of the theory which presumably is very close to the value of $m_A$.

In order to estimate the factors in the above integral we need to consider the gap equation for the gluon field. The term of interest is the quadrilinear gluon interaction one:
\begin{eqnarray}
-\frac{1}{4}g^2f^{abc}f^{ade} A^b_{\mu}A^c_{\nu}A^{d\mu}A^{e\nu}.
\label{int105749}
\end{eqnarray}
We differentiate the expression in Eq. (\ref{int105749}) with respect to the field $A^{m\rho}$ and introduce the gluon condensate to get:
\begin{eqnarray}
9ig^2A^m_{\rho}\int \frac{d^4k}{(2\pi)^4}\frac{1}{k^2-m_A^2}=\frac{9g^2}{16\pi^2}[\Lambda^2-m_A^2\ln[\frac{\Lambda^2+m_A^2}{m_A^2}]]A^m_{\rho}.
\label{res74920uy}
\end{eqnarray}
Then the gap equation determines;
\begin{eqnarray}
m_A^2=\frac{9g^2}{16\pi^2}\Bigg[\Lambda^2-m_A^2\ln[\frac{\Lambda^2+m_A^2}{m_A^2}]\Bigg]
\label{res5464}
\end{eqnarray}
which coincides with the standard results in the literature \cite{Meyers}. For phenomenological reasons we shall consider $m_A^2$ negative. This can be done because $m_A^2=9g^2\Phi_g$ where $\Phi_g$ is the gluon condesate and can be both positive or negative. We make the change $m_A^2\rightarrow-m_A^2$ and rewrite Eqs. (\ref{int65748}) and (\ref{res5464}) as:
\begin{eqnarray}
&&I=\frac{1}{16\pi^2}\Bigg[\ln[1-x]-\frac{1}{1-\frac{1}{x}}\Bigg]
\nonumber\\
&&\frac{16\pi^2}{9g^2}=-\Bigg[x+\ln[1-x]\Bigg],
\label{res66453}
\end{eqnarray}
where $x=\frac{\Lambda^2}{m_A^2}$.

Our goal is to find from the gluon gap equation an estimate for $x$. For that we need to estimate the coupling at which the gluon field gains mass knowing that in order for our approach to work this coupling must be somewhat smaller than the coupling for the quark confinement and chiral symmetry breaking in order to lead to these through a Nambu Jona Lasinio mechanism. To estimate the coupling constant we consider the approach presented in \cite{Jora4} where it is assumed that at the limit between the perturbative and nonperturbative domains important information can be extracted from the Callan Symanzik equations \cite{Callan}, \cite{Callan1}, \cite{Callan2}.  Thus the two point gluon function $G^2(p,g,m)$ must satisfy the equation:
\begin{eqnarray}
[p\frac{\partial}{\partial p}(1-\gamma_m)+2-\beta(g)\frac{\partial}{\partial g}+2\gamma_3]G^2(p,m,g)=0,
\label{csfdg6453}
\end{eqnarray}
where $p$ is the momentum, $\beta(g)$ is the beta function, $\gamma_3$ is the anomalous dimension of the gluon wave function and we work in the background gauge field method where $\gamma_3(g)=-\frac{\beta(g)}{g}$. Moreover $\gamma_m$ is the anomalous dimension of the fermion mass operator and in first order is given by:
\begin{eqnarray}
\gamma_m=-\frac{1}{m}\frac{d m}{d\ln(\mu)}=6\frac{N^2-1}{2N}\frac{g^2}{16\pi^2}.
\label{res66453}
\end{eqnarray}
In first order one can consider $G^2(p,g,m)\approx g^2f(p)$ where $f(p)$ is a function of the momentum and further write:
\begin{eqnarray}
[p\frac{\partial}{\partial p}(1-\gamma_m)+2+4(\beta_0+\beta_1a)a]G^2(p,g,m)=0,
\label{res53423}
\end{eqnarray}
where we denoted:
\begin{eqnarray}
&&a=\frac{g^2}{16\pi^2}
\nonumber\\
&&\beta(a)=-\beta_0a^2-\beta_1a^3,
\label{res629564}
\end{eqnarray}
and one factor of $g^2$ is included in the expression for $G^2(p,g,m)$.
Then one can solve the Callan Symanzik equation to determine that the two point function behaves like:
\begin{eqnarray}
G^2(p,g,m)\approx \frac{1}{p^{\frac{2+4(\beta_0+\beta_1a)a}{1-\gamma_m}}}.
\label{expr77564}
\end{eqnarray}
Next we require that the two point function is of the confining type  $G^2(p,g,m)\approx \frac{1}{p^4}$ and solve the equation:
\begin{eqnarray}
\frac{2+4(\beta_0+\beta_1a)a}{1-\gamma_m}=4
\label{res6453829}
\end{eqnarray}
to determine $\frac{1}{a}=37.4205$. We then solve the gap equation for the parameter $x$ to find out $x=0.994212$ which further leads to $I=1.05513$.

From Eq. (\ref{filre619047}) we calculate the four scalar interaction term as:
\begin{eqnarray}
B=i3g^{\prime4}I \frac{5}{12}\int d^4 x L^{gE\dagger}(x)R^{tE}(x)R^{tF\dagger}(x)L^{gF}(x),
\label{res647856591}
\end{eqnarray}
where we substituted $N=3$. Then the kinetic hamiltonian is diagonalized to lead to the eigenstates:
\begin{eqnarray}
&&S_1=\frac{L+R}{\sqrt{2}}
\nonumber\\
&&S_2=\frac{L-R}{\sqrt{2}}
\label{res74839}
\end{eqnarray}
with the corresponding masses:
\begin{eqnarray}
&&m_1=-Ig^{\prime4}\frac{5}{4}\langle R^{gF\dagger}L^{gF}\rangle
\nonumber\\
&&m_2=Ig^{\prime4}\frac{5}{4}\langle R^{gF\dagger}L^{gF}\rangle,
\label{mass537289}
\end{eqnarray}
where $\langle R^{gF\dagger}L^{gF}\rangle=\langle L^{gF\dagger}R^{gF}\rangle=\alpha'\Lambda'$  where  $\alpha'$ is the tetraquark vacuum condensate and $\Lambda'$ is the corresponding scale. Here we consider the fields $L^{gF}$ and $R^{gF}$ as having mass dimension $1$.
We are interested however only in the absolute values of these masses. We thus consider that the mass of the slightly bound diquark state is just given by the sum of the component quark masses at the scale where these states exist $m_S=2m_q'$. We then use,
\begin{eqnarray}
\frac{1}{2}\gamma_m=-\frac{1}{m}\frac{\partial m}{\partial\ln[\mu^2]}=3\frac{N^2-1}{2N}a=\gamma_0a,
\label{res53929}
\end{eqnarray}
to determine,
\begin{eqnarray}
m_q'=m_q\exp[\gamma_0a\ln[\frac{\Lambda^2}{\Lambda^{\prime 2}}]].
\label{run5649923}
\end{eqnarray}
At this stage we further need to extract the behavior of scales and that of the coupling constants. We shall consider the value of the coupling constant at chiral symmetry breaking $a=\frac{1}{8}$ as reference value that indicates the transition between a region where the beta function has $\beta_0=9$ (see eq, (\ref{coef453789})) and a region where the beta function contain the diquark states and the color triplet baryons which has $\beta_0'=\frac{7}{2}$ (see Eq. (\ref{functr6645})). We start by writing the coupling constant integrated from the beta function for the two regions:
\begin{eqnarray}
&&\frac{1}{a_1}-\frac{1}{a_0}=\beta_0\ln[\frac{\mu_1^2}{\mu_0^2}]
\nonumber\\
&&\frac{1}{a_1}-\frac{1}{a_0'}=\beta_0'\ln[\frac{\mu_1^2}{\mu_0{\prime 2}}].
\label{intd64839}
\end{eqnarray}
Here $\mu_0$ and $\mu_0'$ are the scales where $a_0=\infty$ and $a_0'=\infty$ for the two beta function and $a_1$ is the common value for the coupling at the scale $\mu_1$. In our approach we shall take $\mu_0=\Lambda$ and $\mu_0'=\Lambda'$ (according to the standard picture of low energy QCD where $\Lambda$ is the scale where the strong coupling constant is infinity). From Eq. (\ref{intd64839}) we determine:
\begin{eqnarray}
&&\mu_0^{\prime 2}=\mu_0^2\exp[\frac{1}{a_1\beta_0}-\frac{1}{a_1\beta_0'}]
\nonumber\\
&&\frac{a_0}{a_0'}=\frac{g^2}{g^{\prime 2}}=\frac{\beta_0'}{\beta_0}.
\label{res74536728}
\end{eqnarray}
Substituting the first relation in Eq. (\ref{res74536728}) and the correct values for all the quantities  into Eq. (\ref{run5649923}) further yields:
\begin{eqnarray}
&&m_q'=m_q\exp[\frac{1}{2}\ln[\frac{\Lambda^2}{\Lambda^{\prime 2}}]]= m_q\exp[\frac{44}{63}]
\nonumber\\
&&\Lambda'=\Lambda\exp[-\frac{44}{63}].
\label{res87658}
\end{eqnarray}

Wrapping up all the results in Eqs. (\ref{res55464}),  (\ref{mass537289}) and the subsequent equations we obtain:
\begin{eqnarray}
&&(2m_q')^2=\frac{5}{4}Ig^{\prime 4}\alpha'\Lambda'
\nonumber\\
&&m_q=\frac{3}{4}g^2\alpha\label{res72947},
\label{res4219490}
\end{eqnarray}
which leads to,
\begin{eqnarray}
\alpha'=\frac{4}{5}(\frac{7}{12})^2\frac{1}{I\Lambda}\alpha^2\exp[\frac{3}{2}(\frac{88}{63})]=0.02484 \,\,{\rm GeV}
\label{res82639}
\end{eqnarray}
which is in very close agreement with the results ($\alpha'=0.0249$ GeV) obtained in \cite{Jora5}, \cite{Jora6} from a linear sigma model with two chiral nonets in the limit of an $SU(3)_V$ symmetry after chiral symmetry breaking.

\section{Phase diagram with cascade effect in regular QCD}

We consider QCD with $N_f$ flavors and our purpose here is to determine the number of flavors at which the second confinement and chiral phase transition occur. We start by outlining a picture for the first confinement and chiral symmetry breaking phase transitions \cite{Appelquist1}, \cite{Appelquist2}. It is assumed that confinement of quarks occurs at the infrared fixed point  ($\beta(g)=0$) where the anomalous dimension of the fermion mass operator $\gamma_m=-m\frac{dm}{d\ln(\mu)}=\frac{1}{2}$ \cite{Appelquist2} ($\gamma_m=6\frac{N^2-1}{2N}a$). This would correspond to a coupling constant $a_{01}=\frac{1}{16}$.  Then chiral symmetry breaking with the formation of the two quark condensate takes place for $\gamma_m=1$ and consequently $a_1=\frac{1}{8}$.  The number of flavors at which the confinement phase transition occurs can be calculated easily \cite{Appelquist2}:
\begin{eqnarray}
N_{01f}=\frac{2(-33N+50N^3)}{5(-3+5N^2)},
\label{res71945}
\end{eqnarray}
whereas that of chiral symmetry breaking is \cite{Jora4}:
\begin{eqnarray}
N_{1f}=\frac{-33N+67N^3}{-9+18N^2}.
\label{nr647867}
\end{eqnarray}

The next step is to consider the second confinement and chiral phase transitions where tetraquark and pentaquark states may form. This can be happening after slightly bound diquark states situated in the antitriplet representation of the color group or three quark baryons situated in a triplet of the color group appear in the theory. These states couple with the gluon fields with the coupling $g'$ governed by the beta function $\beta'(a')$ given in Eq. (\ref{functr6645}). Thus at an even lower scale the colored meson and baryon states couple with each other to form singlet tetraquark mesons and pentaquark baryons. We consider the running of the two couplings $a$ and $a'$:
\begin{eqnarray}
&&\frac{1}{a_1}-\frac{1}{a_0}=\beta_0\ln[\frac{\mu_1^2}{\mu_0^2}]
\nonumber\\
&&\frac{1}{a_1'}-\frac{1}{a_0'}=\beta_0'\ln[\frac{\mu_1^{\prime 2}}{\mu_0^{\prime 2}}]
\label{coupl647565}
\end{eqnarray}
Here  $a_1=a_1'=\frac{1}{8}$ as common point at the scale $\mu_1=\mu_1'$ where the two quark condensate forms. Moreover $\mu_0$ and $\mu_0'$ are the scales where the two coupling constants go to infinity $a_0=a_0'\rightarrow\infty$. Then one infers from Eq. (\ref{coupl647565}):
\begin{eqnarray}
\mu_0^{\prime 2}=\mu_0^2\exp[\frac{1}{a_1\beta_0}-\frac{1}{a_1\beta_0'}].
\label{res18350}
\end{eqnarray}
We denote by  $a_1'$ the coupling constant at the scale $\mu_1'$ where the tetraquark condensate forms.  Then one can write:
\begin{eqnarray}
&&\mu_1^{\prime 2}=\mu_0^{\prime2}\exp[\frac{1}{a_1'\beta_0'}]=
\mu_0^2\exp[\frac{1}{a_1'\beta_0'}+\frac{1}{a_1\beta_0}-\frac{1}{a_1\beta_0'}]=
\nonumber\\
&&=\mu_1^2\exp[\frac{1}{a_1'\beta_0'}-\frac{1}{a_1\beta_0'}].
\label{scales647389}
\end{eqnarray}

In order to determine the coupling we will consider this time the two quark and tetraquark vacuum condensates at the scales at which they form, respectively  $\mu_1$ and $\mu_1'$.  By reiterating the procedure in the section III and taking into account that the majority of coefficients and group factors are the same for the two choices of scales one can compute the relevant ratio for the new scales $\mu_1$ and $\mu_1'$.  Here we will give only the final results that relate the two scales pertaining the independence of the vacuum condensates;
\begin{eqnarray}
\frac{\beta_0}{\beta_0'}=\frac{a_1'}{a_1}\exp[\frac{3}{4}\frac{1}{a_1'\beta_0'}-\frac{1}{2}\frac{1}{a_1\beta_0}].
\label{res519370}
\end{eqnarray}

 \begin{figure}
\begin{center}
\epsfxsize = 10cm
\epsfbox{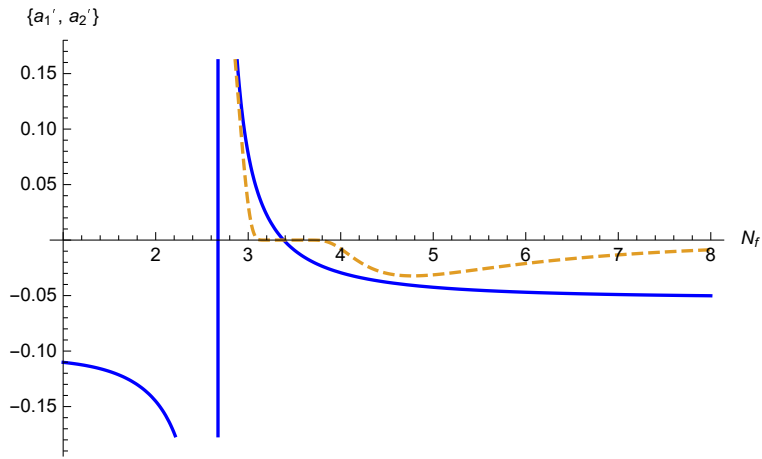}
\end{center}
\caption[]{%
Plot of the quantities $a_1'$ (blue line)  and $a_2'$ (dashed line) as a function of the number of flavors. The points where the two curves intersect ($a_1'=a_2'$) correspond to the critical number of flavors.
}
\label{tetraquark}
\end{figure}

In Fig. \ref{tetraquark} we plot $a_1'=-\frac{\beta_0'}{\beta_1'}$ and $a_2'=\frac{\beta_0a_1}{\beta_0'}\exp[-\frac{3}{4}\frac{1}{a_1'\beta_0'}+\frac{1}{2}\frac{1}{a_1\beta_0}]$ to find the number of flavors that are solutions of the two equations: $\beta'(a')=0$ and Eq. (\ref{res519370}) (where the two curves intersect). Here $N=3$. It is observed that the phase transition  happens for $\beta_0'(a_1')=0$ and for $N_f=3.38$. It turns out that this number also corresponds to the number of flavors below which the asymptotic freedom for $\beta'(a')$ sets in. This result is very interesting because it suggests that for $N_f\geq 4$ it is not possible to have tetraquark condensates.

 In \cite{Jora7} we showed that it is not possible to construct an adequate chiral linear sigma model with tetraquark mesons for $N_f\geq4$ and that in a sense three quark flavors are special. This is because, for example, for $N_f=4$ the tetraquark states are situated in a $(6,\bar{6})$ of $(L,R)$ and this contradicts what we know about chiral symmetry in the context of  three light quark flavors. In our approach we reinforce this  point of view for tetraquark condensates showing that for $N_f\geq4$ the phase diagram excludes the possibility that "four quark" condensates may form.  Thus the standard picture with spontaneous chiral symmetry breaking for a chiral model with three light flavors with both "two quark" and "four quark" states is strengthen. Of course the other heavier flavors may lead to "two quark"  mesons and condensates or "four quark mesons" in a different set-up.

\section{Phase diagram with cascade effect in a  composite model}

We consider a composite picture \cite{Jora1} where at a higher scale there are $N_f'$ fermions situated in the complex conjugate representation of the $SU(3)$ group.  The corresponding beta function at two loops at this scale is $\beta''(a)$  given in Eq. (\ref{thirdbetafunc}).  The first chiral symmetry phase transition should occur for $\gamma_m=6C_2(F)x_1=16x_1=1$  (note that in this case $C_2(F)=\frac{8}{3}$) where we denote $x_1=\frac{g^2}{16\pi^2}$ where g is the strong coupling corresponding to the $SU(3)$ group at higher scales.   Then the chiral symmetry breaking happens at the infrared fixed point where  $\beta''(x_1)=0$ which corresponds to $N_{f1}\approx 1.72$. This result is salutary for our composite model with two flavors because  the formation of a two preon vacuum condensate would break the group $SU(2)_L\times SU(2)_R$ down to $SU(2)_V$ and would contradict any association with the standard model of the composite picture proposed in \cite{Jora1}. However since the number of flavors in our model is larger than the number of  flavors at which chiral symmetry breaking occurs this phenomenon cannot take place and other mechanism should be at play.

In order to find the number of flavors at which the vacuum condensates of four preons form we apply entirely the procedure in section IV to the new couplings and beta functions $\beta''(x)$ and $\beta'''(x)$ from Eqs. (\ref{thirdbetafunc}) and (\ref{res528901}) to find the relation analogous to Eq. (\ref{res519370}) for the preon composites. This reads:
\begin{eqnarray}
x_1'=\frac{\beta_0''x_1}{\beta_0'''}\exp[-\frac{3}{4}\frac{1}{x_1'\beta_0'''}+\frac{1}{2}\frac{1}{x_1\beta_0''}],
\label{preoncom758467}
\end{eqnarray}
where $x_1'=\frac{g_1^{\prime2}}{16\pi^2}$ is the coupling for the second chiral symmetry breaking phase transition.

 \begin{figure}
\begin{center}
\epsfxsize = 10cm
\epsfbox{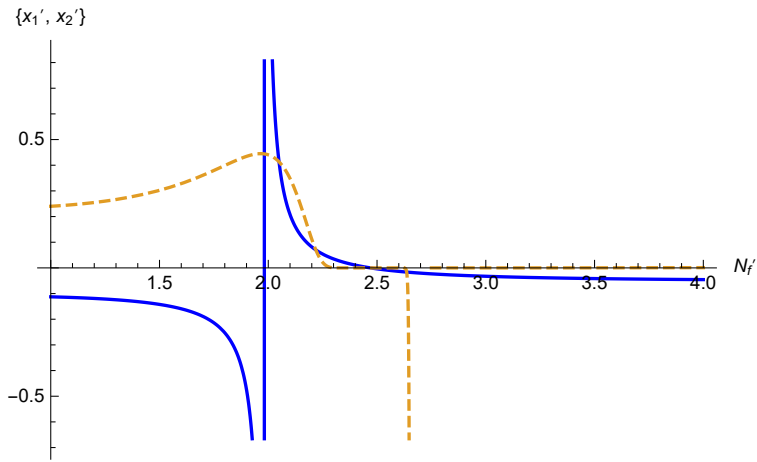}
\end{center}
\caption[]{%
Plot of the quantities $x_1'$  (blue line) and $x_2'$ ( dashed line) as a function of the number of flavors. The points where the two curves intersect ($x_1'=x_2'$)  corresponds to the critical number of flavors.
}
\label{tetrapreon}
\end{figure}

In Fig. \ref{tetrapreon} we plot $x_1'=-\frac{\beta_0'''}{\beta_1'''}$ and $x_2'=\frac{\beta_0''x_1}{\beta_0'''}\exp[-\frac{3}{4}\frac{1}{x_1'\beta_0'''}+\frac{1}{2}\frac{1}{x_1\beta_0''}]$ to find the number of flavors at which second chiral phase transition occurs corresponding to the number of flavors where the two curves intersect.  There are three points of intersection in the region $2\leq N_f'\leq 3$. Any of these points may correspond to the critical number of flavors. However since the number of flavors must be an integer we can only take  the integer part of the corresponding numbers to get $N_f\leq 2$.  This means that the second chiral symmetry breaking sets in for an integer $N_f\leq 2$ suggesting that our composite model displays this phase transition.

The presence of a  vacuum condensate that breaks the electroweak group indicates that the tetrapreon condensates may form  at the electroweak scale so we can set $s'\approx 200$ GeV. We then apply Eq. (\ref{scales647389}) to the two scales of interest for the composite model $s_1'$ and $s_1$ which corresponds to the scale where first confinement and chiral symmetry breaking takes place to find:
\begin{eqnarray}
s_1^2=s_1^{\prime 2}\exp[\frac{1}{x_1\beta_0'''}-\frac{1}{x_1'\beta_0'''}].
\label{sc26178}
\end{eqnarray}
which indicates that the scale of compositeness is around $s_1'\approx 780$ GeV.

\section{Conclusions}

In this work we introduced a toy model to examine the possibility of cascade confinement and chiral symmetry breaking which refers to a phase diagram of a strong group  with fermions in an arbitrary representation $R$  that contains several stages of confinement and chiral symmetry breaking each one corresponding to a different phase transition and to a different number of fermions that bind together or condensate. We discussed in detail two examples: QCD at a lower scale where "four quark" or "five quark" states may form and condensate and QCD at a higher scale in a composite picture.  Based on a simple Nambu Jona Lasinio mechanism we calculated the tetraquark condensate, or more exactly the vacuum condensate of the "four quark" scalars. Our result agrees very well with that obtained from an effective model, a linear sigma model with two chiral nonets, one with a "two quark" structure, the other one with "four quark" mesons \cite{Jora3}.

We analyzed aspects of the phase diagram of QCD in terms of the number of flavors and found out that the second chiral phase transition corresponding to the formation of the "four quark" condensate cannot occur for $N_f\geq 4$ showing  that indeed $N_f=3$ may be magic regarding the tetraquarks structure.

We also considered a hypothetical preon model based on the $SU(3)$ group at higher scales and showed that for $N_f\geq 2$ two preon condensates may not form. However for $N_f\leq2$  "four preon" condensates are allowed and in our picture should correspond to the breaking of the electroweak group. Finally we determined the scale of compositeness as being around $780$ GeV very much within the reach of LHC. However the connection with the LHC experimental results and other phenomenological aspects should be discussed in a future work.

In the end it is useful to stress out the importance of our analysis.  Whereas phase diagram for a non abelian gauge theory with fermions in various representations have been long studied theoretically at zero and finite temperature and through lattice simulations all these studies have disregarded the possibility that multifermion states may correspond themselves to a different state of matter. Tetraquark states have already an established role in phenomenological models  of low energy QCD.  The idea considered also recently in \cite{Pisarski} that the formation of tetraquark condensates indicates a new phase transition is both challenging and intriguing.  In our work we explored the consequences of such an idea in the context of an effective model of the Nambu Jona-Lasinio type. Our results support the picture in which the tetraquark condensates lead to a new phase in  the zero temperature phase diagram but also our previous findings regarding the behavior of tetraquark states in linear sigma models depicting the low energy QCD.  Our conclusion may extend however besides QCD and  have relevant consequences also for  composite models of the electroweak sector of the standard model.

\section*{Acknowledgments} \vskip -.5cm

The work of R. J. was supported by a grant of the Ministry of National Education, CNCS-UEFISCDI, project number PN-II-ID-PCE-2012-4-0078.

\end{document}